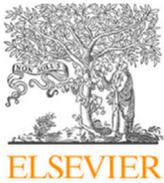
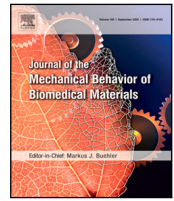

Research paper

# Processing and mechanical properties of novel biodegradable poly-lactic acid/Zn 3D printed scaffolds for application in tissue regeneration

C. Pascual-González [a,b,*], J. de la Vega [a], C. Thompson [a,c], J.P. Fernández-Blázquez [a], D. Herráez-Molinero [a,d], N. Biurrun [a,d], I. Lizarralde [a,d,e], J. Sánchez del Río [a,f], C. González [a,d], J. LLorca [a,d,**]

[a] *IMDEA Materials Institute, C/Eric Kandel 2, 28906 Getafe, Madrid, Spain*
[b] *Materials Science and Engineering Area, Rey Juan Carlos University, C/Tulipán s/n, 28933, Móstoles, Madrid, Spain*
[c] *Department of Material Science and Engineering, Universidad Carlos III de Madrid, 28911 Leganés, Madrid, Spain*
[d] *Department of Materials Science, Polytechnic University of Madrid/Universidad Politécnica de Madrid, E. T. S. de Ingenieros de Caminos, 28040 Madrid, Spain*
[e] *Hexcel Composites S.L., C/ Bruselas 10-16, 28983, Parla, Madrid, Spain*
[f] *Departamento de Ingeniería Eeléctrica, Electrónica Wutomática y Física Aplicada, Universidad Politécnica de Madrid, E. T. S. de Ingeniería y Diseño Industrial, 28012 Madrid, Spain*

ABSTRACT

The feasibility to manufacture scaffolds of poly-lactic acid reinforced with Zn particles by fused filament fabrication is demonstrated for the first time. Filaments of 2.85 mm in diameter of PLA reinforced with different weight fractions of μm-sized Zn - 1 wt.% Mg alloy particles (in the range 3.5 to 17.5 wt.%) were manufactured by a double extrusion method in which standard extrusion is followed by precision extrusion in a filament-maker machine. Filaments with constant diameter, negligible porosity and a homogeneous reinforcement distribution were obtained for Zn weight fractions of up to 10.5%. It was found that the presence of Zn particles led to limited changes in the physico-chemical properties of the PLA that did not affect the window temperature for 3D printing nor the melt flow index. Thus, porous scaffolds could be manufactured by fused filament fabrication at 190 °C with poly-lactic acid/Zn composites containing 3.5 and 7 wt.% of Zn and at 170 °C when the Zn content was 10.5 wt.% with excellent dimensional accuracy and mechanical properties.

## 1. Introduction

Temporary implants and scaffolds from bioresorbable materials, that can be progressively degraded and absorbed in the human body, have tremendous potential for tissue engineering applications (Sheikh et al., 2015; Li et al., 2020; Singh et al., 2020). As opposed to permanent implants, that remain as foreign materials in the body after healing and may require second surgeries to remove the implant because of inflammatory reactions, bioresorbable implants disappear and only the natural tissue remains. Among biodegradable polymers, poly-lactic acid (PLA) is one of the most widely used and it has been approved by the Food and Drug Administration (FDA) of the United States for implantation in the human body for tissue fixation devices and drug-delivery systems (Gunatillake and Adhikari, 2003; Maurus and Kaeding, 2004; da Silva et al., 2018; Chen et al., 2007). Moreover, PLA is a thermoplastic polymer very suitable to be deposited using the fused filament fabrication (FFF) technique (Tyler et al., 2016). Thus, patient-specific implants and scaffolds of PLA can be easily manufactured by additive manufacturing from computer-assisted tomography medical images (Auricchio and Marconi, 2016).

Obviously, the mechanical properties, degradation rate and degradation products of PLA are not optimum for all biomedical applications. For instance, the mechanical properties are low and the degradation times too long (1 to 2 years) for orthopaedic applications (Agüero et al., 2019). Moreover, although products resulting from the biodegradation of PLA are non-toxic and can be metabolized by the body, they contain acidic components that do not always favor tissue integration (Rokkanen et al., 2000). In order to overcome these limitations, composites made up of a PLA matrix reinforced with metallic particles have been manufactured. For instance, PLA/Ti composite filaments, suitable for FFF, were prepared by a double thermal mixing route. The tensile and compressive mechanical properties as well as the cytocompatibility of porous scaffolds manufactured by FFF were enhanced when the volume fraction of Ti particles was ≤ 10%. Nevertheless, this work did not provide evidence that high quality filaments (with constant diameter, good particle dispersion and negligible porosity) could be manufactured by this route and, in addition, Ti particles are inert but not biodegradable. Similar results and limitations were reported in the case of PLA filaments reinforced with 316L stainless steel particles (Jiang and Ning, 2020).

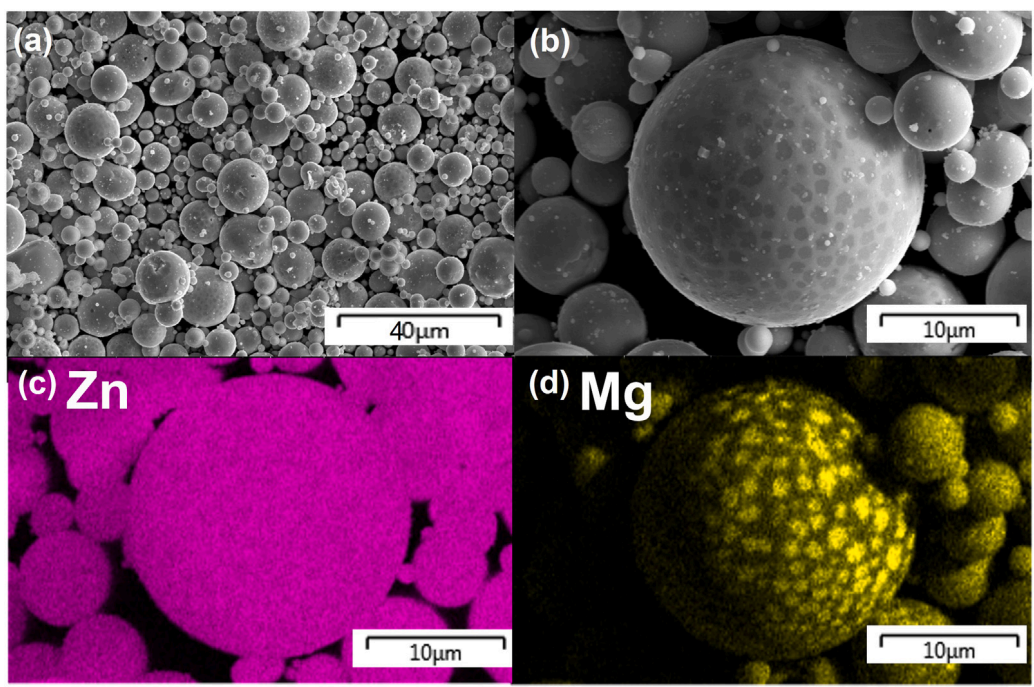

**Fig. 1.** (a) and (b) Secondary electrons SEM images of the Zn-Mg1 powder. (c) and (d) EDS analysis showing the distribution of Zn and Mg in the particles.

Most of the investigations to reinforce PLA with metallic particles have been focused in Mg that is also biodegradable. Antoniac et al. (2019) reported the viability of PLA/Mg filaments (4 wt% of Mg) for FFF manufactured by conventional extrusion while a colloidal route was developed by Ferrández-Montero et al. (2019) and Ferrández-Montero et al. (2020) that allows higher particle volume fractions. More recently, PLA/Mg filaments were manufactured by a double extrusion method (a standard extrusion followed by a precision extrusion) that allows to produce high quality filaments with constant diameter, full density and homogeneous dispersion of the Mg particles (Pascual-González et al., 2022). They were successfully used to print face-centered cubic scaffolds with optimum dimensional accuracy and fully-dense struts. It should be noted, however, that the combination of PLA and Mg particles has several limitations. Firstly, thermal degradation of PLA at the extrusion and printing temperatures can be catalyzed by the presence of MgO (Feliu Jr. et al., 2009; Esmaily et al., 2017; Motoyama et al., 2007). As a result, the molecular weight (MW), viscosity and mechanical properties of PLA/Mg composites are reduced (Cifuentes et al., 2017) and it is more difficult to find the appropriate processing window during FFF (Pascual-González et al., 2022). Secondly, Mg particles suffer from rapid corrosion rates, especially in chloride solutions such as body fluids, that cause the formation of hydrogen gas and a high local pH around the implant, which may lead to necrosis of tissues (Echeverry-Rendón et al., 2018; Kopp et al., 2019; Li et al., 2021). Finally, the increase in volume of the Mg fibers and particles due to the formation of $Mg(OH)_2$ leads to the cracking of the PLA matrix and to a large reduction in the mechanical properties (Ali et al., 2021).

An interesting option to overcome the problems associated with the high reactivity of Mg is to use another biodegradable metal such as Zn. Zn is an essential element for the body and is involved in a high number of enzymatic reactions in which it participates as a cofactor of more than 300 enzymes (Fraga, 2005). This element is also associated with biological functions that include anti-inflammatory and anti-proliferative properties, as well as in the prevention of endothelial cell apoptosis, atherosclerosis risk reduction, among other beneficial properties to the body (Hennig et al., 1999). Moreover, corrosion rates of Zn are much lower than those of Mg and Zn alloys have demonstrated excellent biocompatibility (Murni et al., 2015; Shen et al., 2016; Gong et al., 2015). However, to the authors' knowledge, the development of PLA/Zn filaments for FFF has not been previously achieved and this is the main objective of this investigation. Dense, homogeneous and printable PLA/Zn filaments with constant diameter were manufactured containing different weight fractions of Zn particles. The influence of the Zn particles and of the processing on the physico-chemical and mechanical properties of PLA/Zn composites was investigated. Finally, porous scaffolds were manufactured by FFF and their microstructure and mechanical properties were analyzed. These results show the viability of the fabrication of PLA/Zn composite scaffolds for applications in regenerative tissue engineering.

## 2. Methods

### 2.1. Materials and filament fabrication

PLA/Zn filaments were manufactured from Ingeo Biopolymer 2003D PLA (NatureWorks LLC, Minnesota, USA) and spherical particles of Zn - 1 wt% Mg (Zn-Mg1) alloy produced by gas atomization by Meotec GmbH (Aachen, Germany). The size, morphology and element distribution in the Zn-Mg1 particles were examined by scanning electron microscopy (SEM) (Helios NanoLab 600i, FEI) using secondary electrons as well as energy dispersive spectrometry (EDS). The PLA pellets were dried in a conventional oven at 60 °C for 24 h and mixed with the Zn alloy powders by means of manual agitation, ensuring that most of the metallic powders were stuck on the pellet surface. PLA and PLA mixed with 3.5, 7, 10.5, 14 and 17.5 wt.% of Zn-Mg1 particles was used. The mixtures were subjected to standard extrusion at 180 °C in a microcompounder (Xplore MC 15) that has two conical mixing screws.

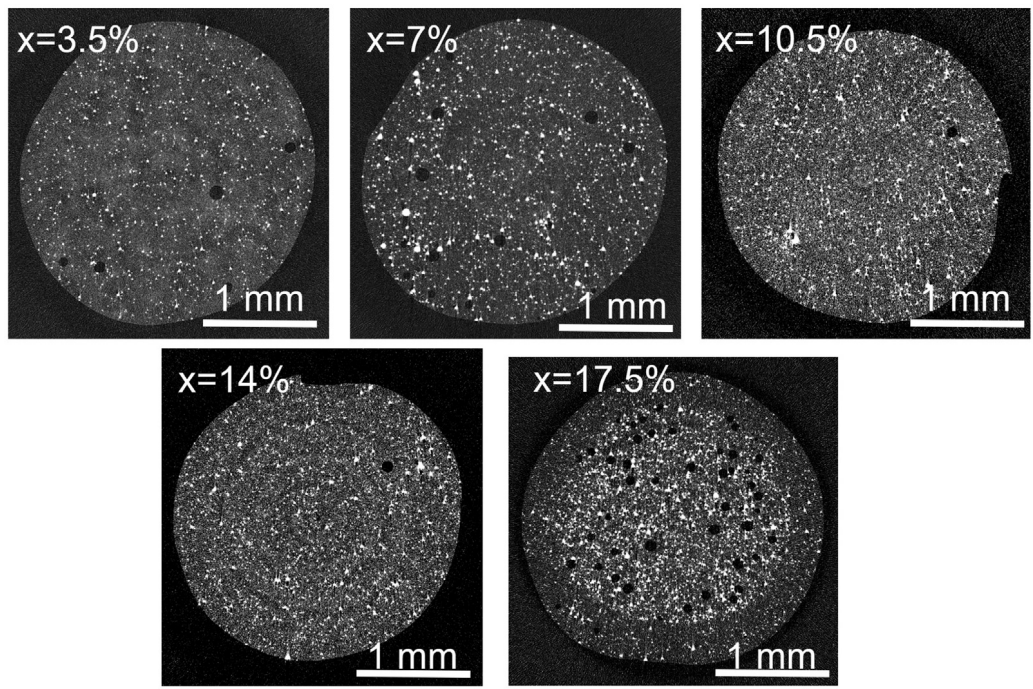

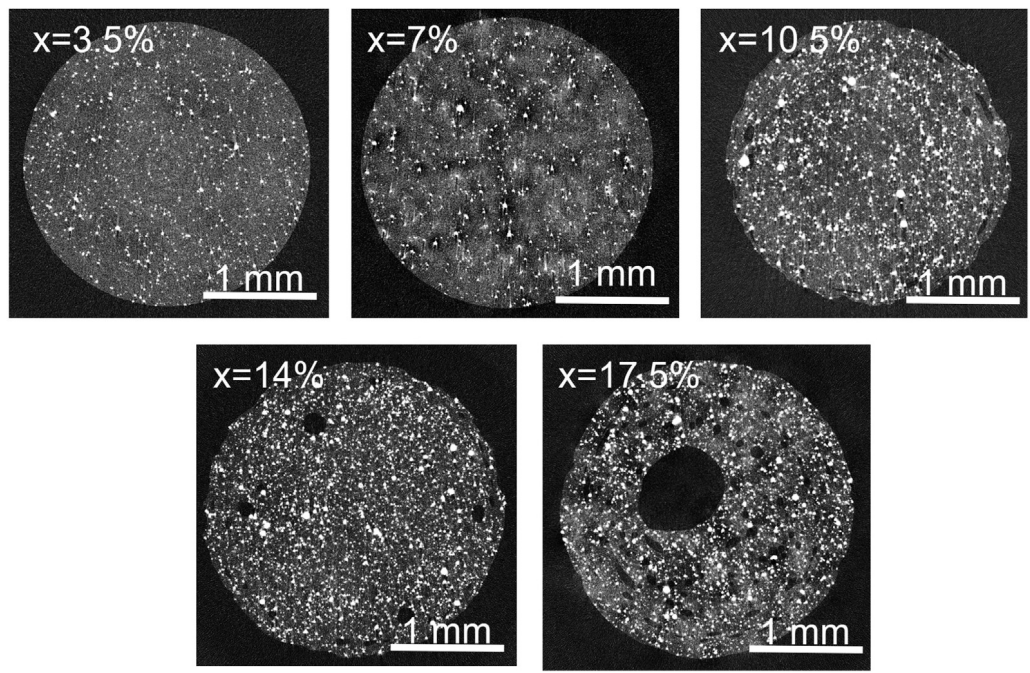

**Fig. 2.** Representative cross sections obtained by XCT of PLA-x Zn (x in wt.%) filaments after (a) standard extrusion and (b) precision extrusion. *x* stands for the wt.% of Zn. Filament diameter was 2.85 mm in all cases. White regions stand for the Zn particle while black regions are pores.

The residence time was approximately 3 min and the rotating speed of the screws was 100 rpm. The diameter of extruded filament was not constant (and, thus, not suitable for 3D printing by FFF). The PLA/Zn filaments were pelletized (Granulator, Brabender) and the composite pellets were used to feed a filament-maker machine that carries out a precision extrusion (Precision 450, 3Devo, Utrecht, The Netherlands). Precision 450 filament-maker includes four heaters along the mixing screw and, in addition, the diameter of the filament is continuously measured with an optical sensor that also provides a feedback control to the automated spooling system. The nominal diameter was set to 2.85 mm, the standard diameter used by the Ultimaker S5 3D printer. The heater temperatures were optimized depending on the metallic particle content and they are indicated in Table 1. The temperature of all heaters was initially 10% above the melting temperature determined by DSC. Afterwards, the temperature of each heater was manually modified until a proper flow of material was attained. This process was guided by the MFI results which indicated that the viscosity decreased with increasing Zn content. The residence time was approximately 12 min. More details about the filament production can be found in Pascual-González et al. (2022).

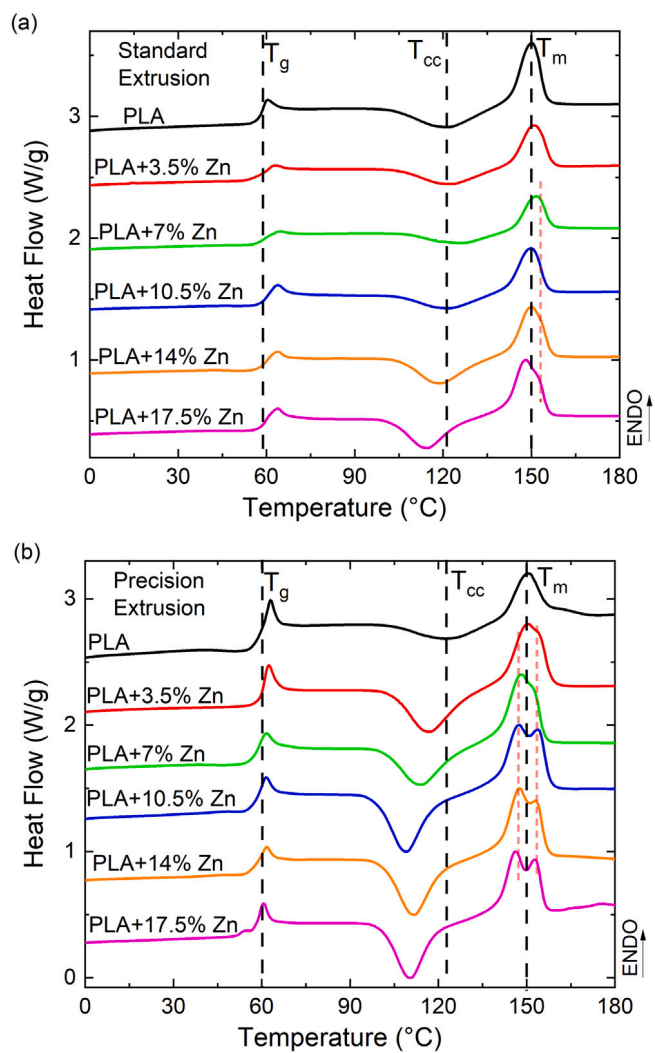

**Fig. 3.** DSC curves during the heating cycle of PLA/Zn filaments at a scanning rate of 10 °C/min. Glass transition ($T_g$), cold crystallization ($T_{cc}$) and melting temperature ($T_m$) of pure PLA are marked with black dashed lines for reference. New emerging peaks in the PLA/Zn composites are indicated with short red dashed lines.

*2.2. Filament characterization*

The internal features of the composite filaments (porosity and particle distribution) were analyzed by X-ray computed tomography (XCT) in a Phoenix Nanotom (General Electric). XCT was carried out with a W target and 0-mode nanofocus at 50 KV with a current of 300 $\mu$A and the voxel size was 2 $\mu$m. In addition, PLA/Zn composite filaments were analyzed by differential scanning calorimetry (DSC) (Q200, TA Instruments) up to 180 °C with a heating rate of 10 °C/min. Thermogravimetric analysis (TGA) (Q50, TA Instruments) was used to evaluate the thermal degradation of PLA/Zn specimens up to 500 °C in air from the weight variation as a function of temperature. The molecular weight (MW) distribution of the PLA/Zn composites was measured by gel permeation chromatography (GPC) (GPC 2414, Waters). The system uses a Waters 2424 refractive index detector and a series of narrow polystyrene standards with tetrahydrofuran as the mobile phase. Melt flow index (MFI) tests of the filament after standard and precision extrusion were carried out in a MFI tester (KJ-3092, Kinsgeo) with a load of 2.16 kg for 10 min following ASTM standard D1238. Finally, variable-temperature X-ray scattering experiments were performed in an Empyrean PANalytical X-ray diffractometer with Ni-filtered Cu K$_\alpha$ radiation. The heating stage is a homemade instrument using a coil heater provided by Resistencias Industriales Maxiwatt and controlled with an Eurotherm 2416 temperature controller.

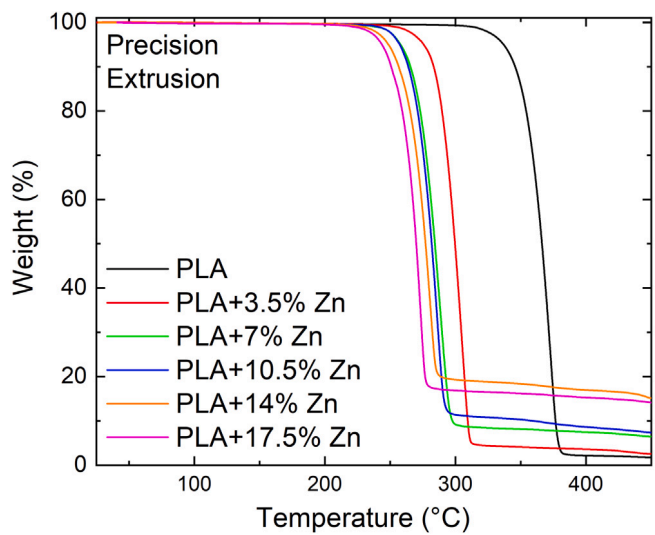

**Fig. 4.** Weight loss as a function of temperature during the TGA test in air.

**Table 1**
Temperature (in °C) of the heaters in the Precision 450 filament-maker during precision extrusion of PLA/Zn composites. Zn content is expressed in wt.%.

| Material | Heater 1 | Heater 2 | Heater 3 | Heater 4 |
|---|---|---|---|---|
| PLA and PLA/Zn (3.5%, 7%) | 180 | 200 | 180 | 170 |
| PLA/Zn (10.5%, 14%) | 170 | 190 | 170 | 160 |
| PLA/Zn (17.5%) | 160 | 180 | 160 | 150 |

**Table 2**
Pore volume fraction (in %) in PLA/Zn filaments processed by standard and precision extrusion.

| Zn content (wt.%) | Standard extrusion | Precision extrusion |
|---|---|---|
| 3.5% | 0.20 | 0 |
| 7% | 0.80 | 0 |
| 10.5% | 0.15 | 0.68 |
| 14% | 0.16 | 2.23 |
| 17.5% | 1.23 | 13.80 |

*2.3. Scaffold 3D printing and characterization*

Scaffolds of 20 × 20 × 20 mm$^3$ following a face-centered cubic (FCC) lattice pattern were 3D printed by FFF with an Ultimaker S5 3D printer with the precision extruded filaments. The struts of the lattice have a circular section of 1.5 mm in diameter. The scaffolds were printed layer-by-layer with a travel velocity of 5 mm/s and a nozzle of 400 $\mu$m in diameter. The printing temperature was in the range 190 °C to 210 °C depending on the PLA/Zn filament composition while the bed temperature was kept at 70 °C. The microstructure of selected scaffolds was studied using XCT. The sample and detector positions were modified to attain a voxel size of 12 $\mu$m. In addition, the elastic modulus and strength of the scaffolds were measured by means of compression tests between under displacement control at crosshead speed of 0.5 mm/min in an Instron 3384 universal mechanical testing machine. Teflon paste was inserted between the scaffold and the compression plates to minimize friction. The applied load was measured with a load cell while the reduction in length was determined by means of a linear variable differential transducer (LVDT) in contact with the compression plates.

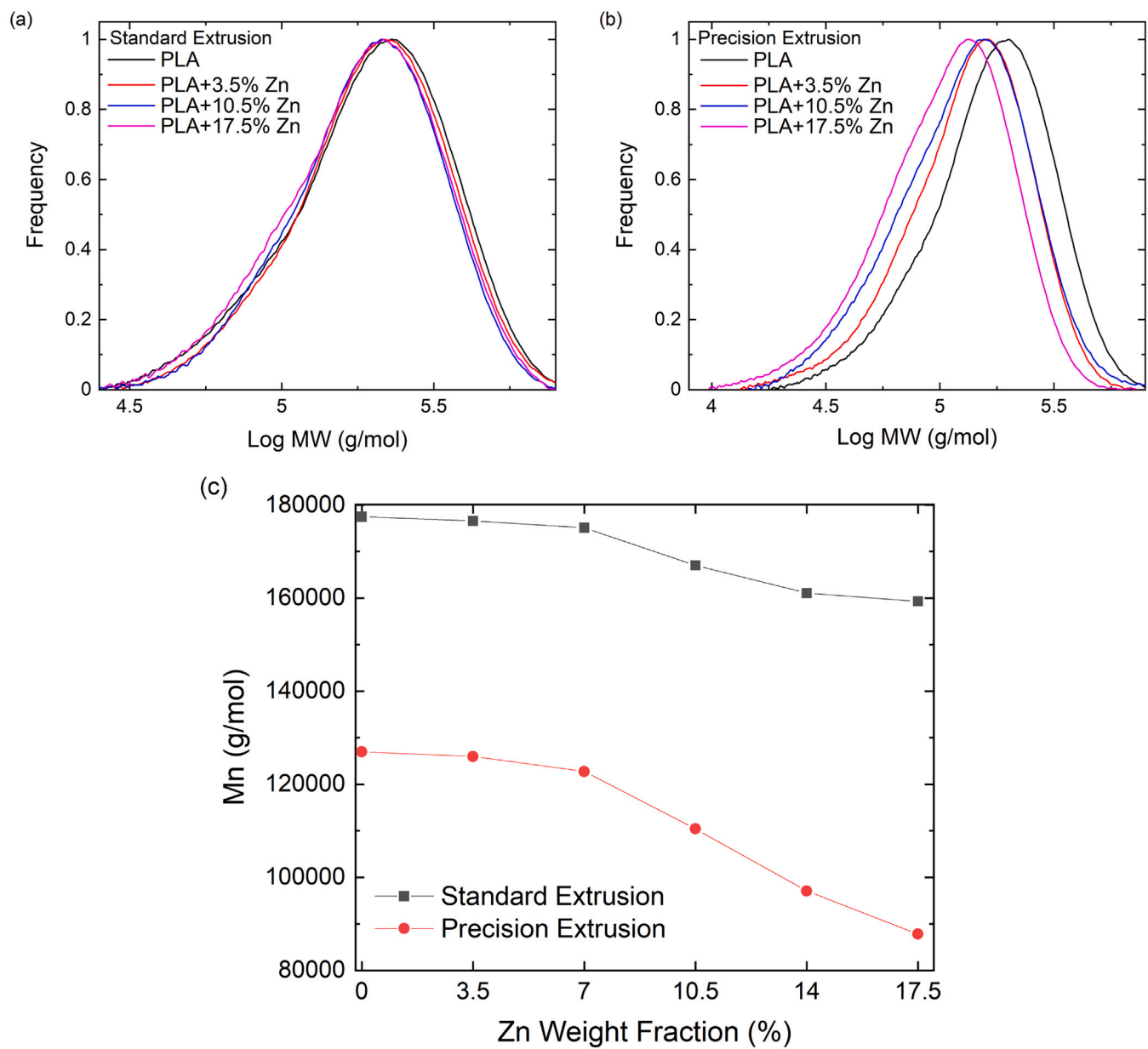

**Fig. 5.** MW distributions of PLA/Zn composites. (a) After standard extrusion. (b) After precision extrusion. (c) Evolution of the number average molecular weight (Mn) with Zn content after standard and precision extrusion.

## 3. Results and discussion

### 3.1. Porosity and particle distribution in PLA/Zn filaments

SEM images of the ZnMg1 particles are shown in Figs. 1 (a) and (b). Particles are spherical and the diameter distribution is in the range 2 μm to 25 μm. The Zn and Mg distribution in the particles was analyzed by EDS and the results are shown in Figs. 1(c) and (d). Zn is homogeneously distributed in the particles which show some Mg-rich regions. The solubility of Mg in Zn is very low and the Mg-rich regions are formed by the $Mg_2Zn_{11}$ eutectic phase (Liu et al., 2020).

The filaments with different weight fractions of Zn (x = 3.5%, 7%, 10.5%, 14% and 17.5%) manufactured by standard extrusion and standard extrusion followed by precision extrusion were analyzed by XCT. Representative cross-sections are depicted in Fig. 2(a) and (b) for standard and precision extruded filaments, respectively. Black regions correspond to pores while white regions stand for spherical Zn particles. In general, a homogeneous distribution of the Zn particles in the PLA matrix is obtained after standard or precision extrusion up to 14 wt.% of Zn. However, the concentration of the filler Zn particles is higher near the center of the filament in the standard extruded filaments with 17.5 wt.% Zn. Filler migration towards the center of the filament during polymer extrusion has been previously reported (Colón Quintana et al., 2019; Goris and Osswald, 2018) and it is the result of the radial forces that develop on the filler particles during extrusion.

The tomograms of filaments of 3 mm in length were segmented to determine the pore volume fraction that can be found in Table 2. White regions stand for the Zn particle while black regions are pores. Standard extruded materials present non-constant diameter and irregular cross-section shape. The porosity was low except for the filaments with 17.5 wt.% Zn (Fig. 2(a)). Precision extruded filaments have constant diameter and circular shape and the porosity of the filaments is practically reduced to zero after precision extrusion for Zn content up to 10.5% (Fig. 2(b)). However, the porosity increases rapidly for 14 wt.% and 17.5 wt.% of Zn particles and large pores can be observed in the filament containing 17.5 wt.% Zn in Fig. 2(b)).

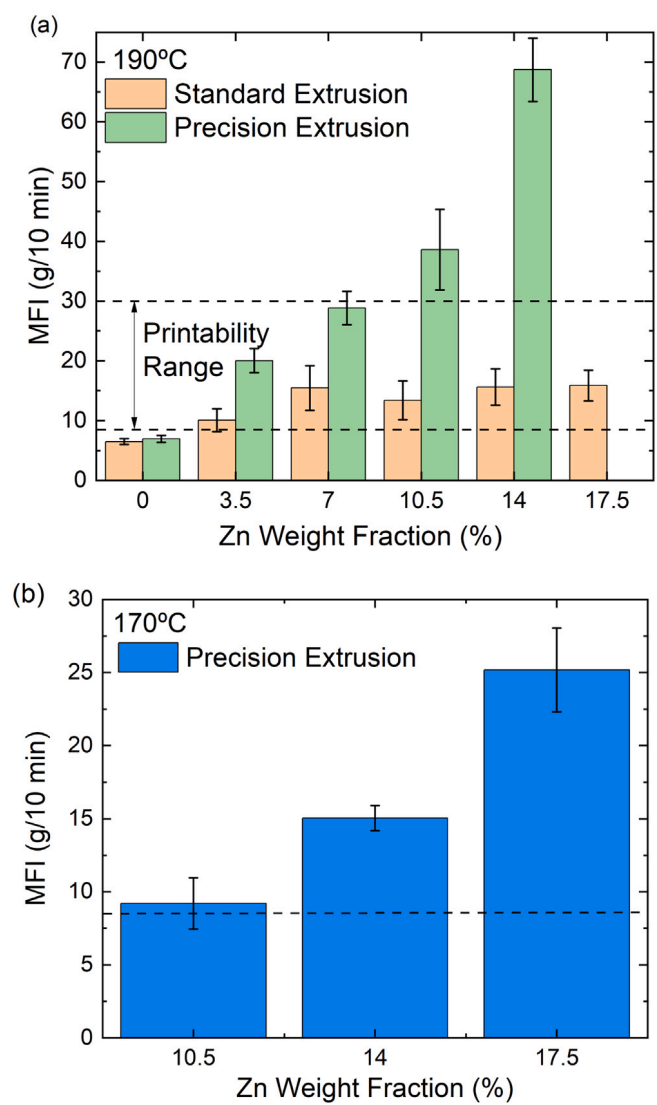

**Fig. 6.** (a) MFI at 190 °C of standard extruded and precision extruded PLA/Zn filaments as a function of the Zn content. (b) MFI at 170 °C of precision extruded filaments of PLA/10.5 wt.% Zn, PLA/14 wt.% Zn and PLA/17.5 wt.% Zn. The MFI printability range is marked by the horizontal dashed lines.

The porosity of the PLA/Zn filaments is higher than that reported in PLA/Mg filaments processed by the same route and with equivalent volume fraction of metallic particles (Pascual-González et al., 2022). The higher porosity of the PLA/Zn filaments may be attributed to the large content of small Zn particles (< 5 μm), which increases the contact surface between metal and polymer. This phenomenon leads to higher porosity because of the gas layers surrounding the particles (Goris and Osswald, 2018) and of the higher effective viscosity of suspension that results in a higher gas volumes trapped in the melt as well as improper filling of the gaps between adjacent particles (Colón Quintana et al., 2019). In addition, the air trapped within particle clusters and the limited metal flow inside them can contribute to the formation of pores (Colón Quintana et al., 2019). In particular, the pore volume fraction in the filaments increased rapidly when the Zn content was around 14%, particularly in the precision extruded filaments. This latter behavior may be attributed to the particular processing conditions associated with precision extrusion.

Overall, it should be noted that standard extrusion followed by precision extrusion is able to produce continuous, uniform and dense PLA/Zn filaments which are suitable for FFF. Furthermore, this process is reproducible, predictable and can be easily scaled up. The porosity in the filaments increased when the Zn content was around 14 wt.% and increased more rapidly for the precision extruded filament because of the differences in the processing conditions.

*3.2. Physico-chemical properties of PLA /Zn filaments*

DSC curves during heating at 10 °C/min of the PLA/Zn filaments after standard and precision extrusion are plotted in Figs. 3(a) and (b), respectively. All DSC curves first exhibit a peak corresponding to the glass transition ($T_g$), followed by an exothermic phase transition due to cold crystallization ($T_{cc}$), and finally by an endothermic transition due to polymer melting ($T_m$). The glass transition of neat PLA occurs at approximately 58 °C. The $T_g$ of PLA/Zn composites after standard and precision extrusion remains unaffected regardless of the metal content (dashed line at lowest temperatures on Figs. 3(a) and (b)), indicating that polymer chain mobility is not modified by the addition of Zn particles. On the contrary, the metal particles favor PLA crystallization, that occurs at lower temperatures and with narrower exothermic peaks, evidencing the nucleating effect of Zn. The nucleation of PLA starts at ~90 °C and the cold crystallization peak ($T_{cc}$) is reached at ~ 122 °C (dashed lines in Figs. 3(a) and (b)). $T_{cc}$ decreases with increasing Zn content and it reaches ~114 °C after standard extrusion and ~ 110 °C after precision extrusion in the filaments with 17.5 wt.% Zn. The difference in $T_{cc}$ between standard and precision extrusion may be ascribed to the reduction of the MW and of the physical entanglements in PLA due to the shear stress during the precision extrusion, since this second processing step was carried out at higher temperatures and with longer residence times than standard extrusion. Cold crystallization has also been reported to be facilitated by the addition of metal particles in PLA/Mg composites (Antoniac et al., 2019). The melting peak ($T_m$) of PLA after standard and precision extrusion is found at 150 °C (dashed lines in Figs. 3(a) and (b)). $T_m$ of PLA/Zn composites after standard extrusion remains unaffected up to 14 wt% of Zn, and a slight decrease (together with the emergence of a shoulder at higher temperature) is observed in PLA-17.5 wt.% Zn. $T_m$ decreases slightly after precision extrusion due to the reduction in $T_{cc}$ and new endothermic peaks appear at higher temperatures in PLA/Zn composites, which can be associated to melt/recrystallization processes (marked by red short dashed lines on Figs. 3(a) and (b)). The split in the melting peak is more noticeable with increasing Zn content.

The weight loss as a function of temperature during the TGA in air is plotted in Fig. 4 for the PLA/Zn filaments subjected to standard and precision extrusion. PLA degradation begins at ~ 300 °C and finishes at ~390 °C. PLA/Zn composites start to degrade at lower temperatures (220–240 °C) and the end of the degradation also occurs at lower temperatures (275–310 °C). Both temperatures decrease with increasing the Zn content but this effect is minor as compared with presence of relative small amounts of Zn particles. The residual mass increased upon Zn content, except for PLA+14 wt.%, that was slightly higher than that of PLA + 17.5 wt.%. This difference may be attributed differences in the actual Zn content due to the small size of the samples (between 10–15 mg).

The MW distributions of the PLA/Zn filaments after standard and precision extrusion were obtained from the GPC experiments and they are plotted in Fig. 5(a) and (b), respectively. The data from the as-received PLA pellets are added in both figures. The presence of Zn particles led to an almost negligible reduction in the MW distribution of the PLA after standard extrusion, which took place at 180 °C with a residence time of 3 min. The reduction in MW was slightly more noticeable after precision extrusion, which was carried out at a maximum temperature 200 °C with a residence time of 12 min. Thus, the presence of Zn particles induced some scission of the molecular chains during extrusion which increases with the Zn content, as shown in the plot of the average MW (Mn) of PLA/Zn filaments with Zn content in Fig. 5(c). This process was slightly accelerated during precision extrusion because of the higher temperature and residence time. These results are in contrast with those reported for PLA/Mg filaments processed using the same strategy (Pascual-González et al., 2022). The average MW (Mn) of the PLA in PLA/5 wt.% Mg composites decreased more than 50% after precision extrusion at 190 °C and the differences between PLA/Zn and PLA/Mg composites have to be attributed to the thermal degradation of the polymeric chains catalyzed by the presence of the Mg particles.

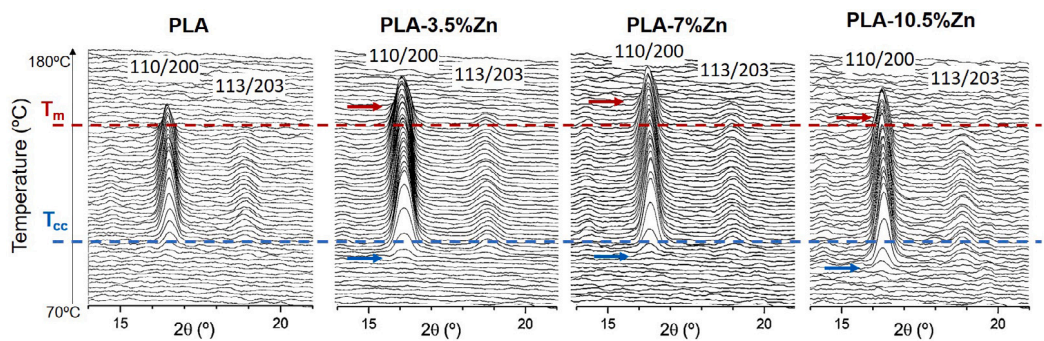

**Fig. 7.** Temperature dependence of XRD patterns of PLA and PLA/Zn composites from 70 °C up to 180 °C.

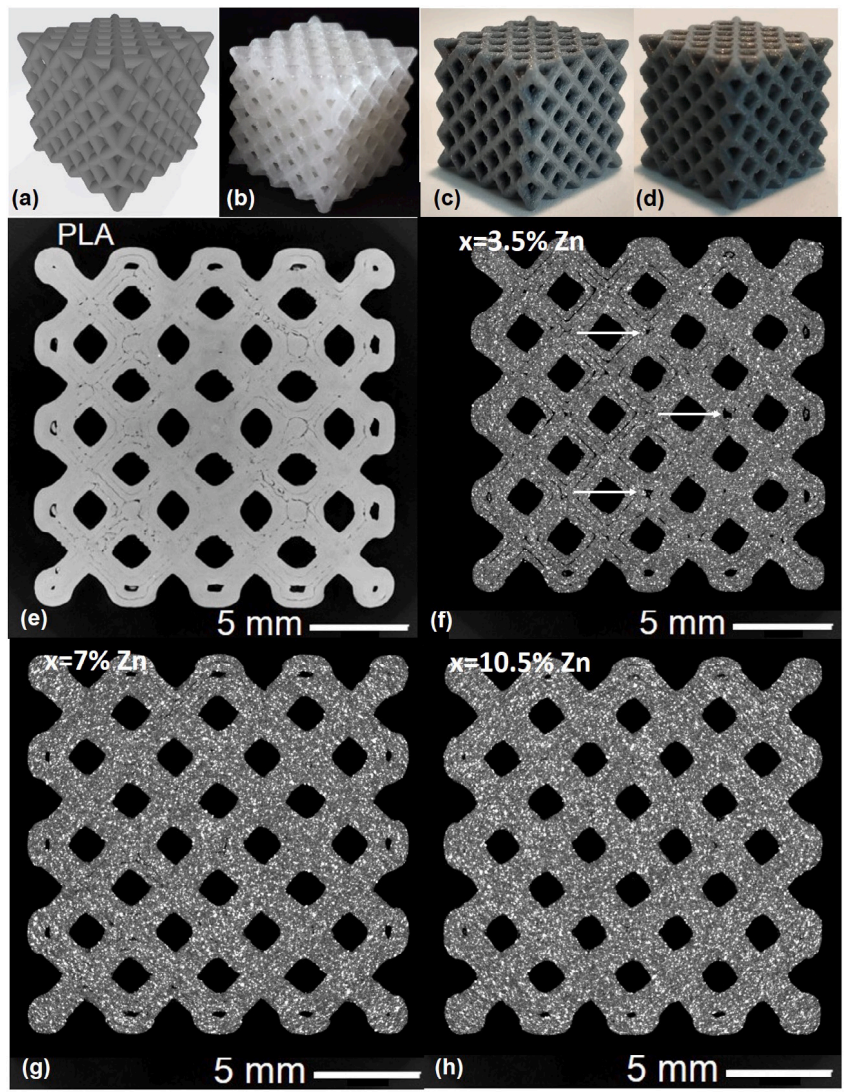

**Fig. 8.** FCC lattice scaffolds of PLA/Zn particles manufactured by FFF. (a) CAD model. (b) Scaffold of PLA, (c) PLA-3.5% Zn and (d) PLA-10.5% Zn. (e)–(h) Cross-sections of FCC scaffolds extracted from XCT tomograms. Internal pores are marked with white arrows.

The MFI is a common test performed to investigate the melt-processability of polymers and to determine the optimum temperature for 3D printing. According to the literature, the MFI of the filament should be in the range of 10 g/10 min to 30 g/10 min to be successfully printed by the FFF technique (Wang et al., 2018; Pascual-González et al., 2022). The MFI of the PLA/Zn composites at 190 °C after standard and precision extrusion is plotted in Fig. 6(a). These results are the average of five tests and the error bars stand for the standard deviation. The MFI of precision extruded filaments increased slightly with the incorporation of up to 7 wt.% of Zn and remained approximately constant (15 g/10 min) for larger Zn contents. These results are in agreement with the limited reduction in the average MW of PLA after standard extrusion, Fig. 5(c), which did not significantly influence the viscosity of the composite. However, the MFI after precision extrusion increased rapidly with the Zn content due to the reduction of the average MW of PLA. As a result, composites with Zn content above 10.5 wt.% were outside of the printability range at 190 °C and it was not possible to measure the MFI of the PLA/ 17.5 wt.% Zn composite because of its high fluidity. Thus, the MFI of the composites with more than 10 wt.% of Zn were determined at 170 °C and the results are plotted in Fig. 6(b). They were within the printability range at this temperature.

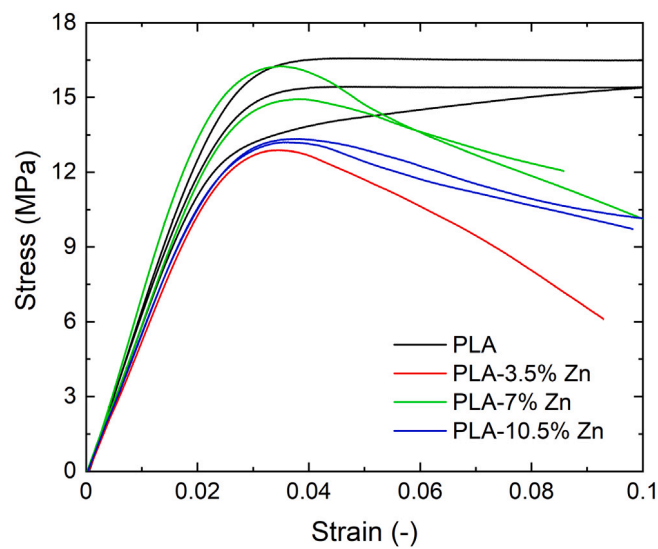

**Fig. 9.** Engineering stress–strain curves in compression of 3D printed scaffolds of PLA and PLA with 3.5, 7.0 and 10.5 wt.% of Zn.

**Table 3**
Mechanical properties of the 3D printed FCC scaffolds deformed in compression.

|  | Elastic modulus (MPa) | Strength (MPa) |
|---|---|---|
| PLA | 650 ± 32 | 16 ± 1 |
| PLA-3.5% Zn | 529 ± 1 | 13 ± 1 |
| PLA-7% Zn | 675 ± 103 | 16 ± 1 |
| PLA- 10.5% Zn | 517 ± 10 | 13 ± 1 |

Overall, the results of the TGA tests (Fig. 4) and of the MFI (Fig. 6) indicate that the presence of Zn particles accelerates the degradation of the PLA. However, the temperature window required for FFF processing – which is found between $T_m$ and the temperature at which the material begins to degrade (220–240 °C) – is not compromised.

The results of DSC (Fig. 3) indicate that Zn particles promote the formation of crystalline nuclei in PLA at lower temperatures. It is interesting to analyze the evolution of the crystalline phase in PLA during heating from room temperature because this strategy might be used to modify the properties of PLA (and, thus, of the composite) in 3D printed scaffolds. The evolution of the XRD patterns of neat PLA and PLA/Zn composites with 3.5, 7 and 10.5 wt.% of Zn from 70 °C up to 180 °C with a heating rate of 10 °C/min is plotted in Fig. 7. At low temperatures (from 70 °C to $T_{cc}$), PLA and PLA/Zn composite do not show diffraction peaks, in agreement with the amorphous glass state that was also observed by DSC (Fig. 3). A reference dashed blue line is added at 120 °C in Fig. 7, corresponding with the cold crystallization of PLA. This value is determined with the emergence of the two main diffraction peaks, corresponding to two doublet forms (100)/(200) and (113)/(203). As indicated by the blue arrows in Fig. 7, the diffraction peaks for PLA-3.5% Zn, PLA-7% Zn and PLA-10.5% Zn appear at lower temperatures than those for PLA, especially for PLA-10.5%Zn composite. This means that lower temperatures are required for the cold crystallization with increasing of Zn content. Moreover, a reference dashed red line, that corresponds to the melting point of PLA, is added at 150 °C in Fig. 7. As indicated by the red arrows in Fig. 7, higher melting temperatures were found in the composites, in agreement with the disappearance of the diffraction peaks.

*3.3. Manufacturing and mechanical properties of PLA/Zn scaffolds*

Porous scaffolds of 20 × 20 × 20 mm with a FCC lattice and circular struts of 1.5 mm in diameter were manufactured by FFF following the CAD model in Fig. 8(a) with precision extruded filaments of PLA, PLA/3.5 wt.% Zn, PLA/7 wt.% Zn and PLA/10.5 wt.% Zn. There is no direct relation between the porosity of the filaments and the scaffolds. The scaffolds were 3D printed using filaments of pure PLA, PLA+3.5% Zn, PLA+7% Zn and PLA+10.5% Zn with porosities below 0.7%. They are shown in Figs. 8(b)–(d), respectively. The printing temperatures were selected according the MFI experiments (Fig. 6): 210 °C for neat PLA, 190 °C for PLA/3.5 wt.% Zn and 170 °C for the composites with 7 and 10.5 wt.% of Zn. It was not possible to print scaffolds with higher Zn content due to clogging problems.

Cross-sections of the XCT tomograms of the scaffolds made up of PLA and PLA/Zn composites with 3.5, 7 and 10.5 wt% Zn are shown in Figs. 8(e)–(h). The figures indicate that the scaffolds have good geometrical accuracy and that the Zn particles were homogeneously distributed in the case of the composites. Moreover, slightly higher internal porosity was observed in the case of the PLA/3.5 wt.% scaffold (marked with arrows in Fig. 8(f)). The addition of higher loads of Zn to the filaments reduced the internal porosity of the scaffolds (and it was possible to print the scaffolds at lower temperatures) probably due to the reduction in viscosity during processing which facilitated the deposition of the material.

The nominal stress–strain curves in compression of the 3D printed scaffolds are plotted in Fig. 9. Stress was calculated from the applied load divided by the initial cross-section of the scaffold while the engineering strain was obtained from the displacement measured by the LVDT divided by the initial height. All the samples were deformed up to 10% compressive strain. The PLA scaffolds showed continuous hardening during deformation (or, at least, a constant flow stress), which is indicative of homogeneous deformation. In fact, the optical micrograph of the lateral section of the PLA scaffold at the beginning of the compression test (Fig. 10(a)) and after 10% of compressive strain (Fig. 10(b)) show a slight barreling of the scaffold after deformation. On the contrary, softening after the peak load was found in all of the scaffolds containing Zn particles, which is indicative of the early localization of damage. This is shown in Fig. 10(d), where the lateral view of the PLA + 7 wt.% Zn scaffold is depicted after 10% compressive strain. Damage has been localized in the first row of the scaffold lattice and vertical cracks (parallel to the loading direction) have developed. Thus, the presence of Zn particles embrittles the material, leading to early crack formation.

The average value and standard deviation of the elastic modulus and of the strength of the scaffolds without and with Zn particles are indicated in Table 3. The mechanical properties of the PLA and PLA + 7 wt.% Zn scaffolds were very close while those of the scaffolds with 3.5 wt.% and 10.5 wt.% Zn were slightly lower, very likely due to the porosity in the former and to the brittleness induced by the Zn particles in the latter. Thus, PLA + Zn particle scaffolds present similar properties to those found in PLA scaffolds and the main objective of the Zn particle reinforcement should be to tailor the degradation rate and the biocompatibility. Both issues will be addressed in the future.

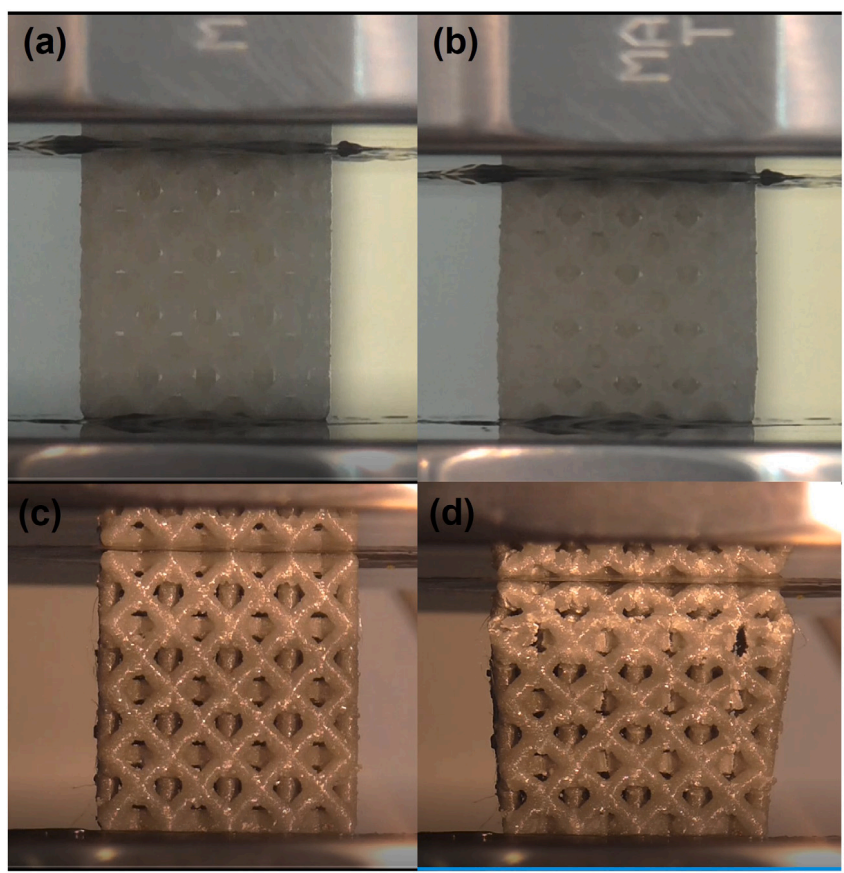

**Fig. 10.** Lateral view of the scaffolds. (a) PLA, before testing. (b) PLA, after 10% compressive strain. (c) PLA + 7 wt.% Zn, before testing. (d) PLA + 7 wt.% Zn, after 10% compressive strain.

### 4. Conclusions

The feasibility to manufacture scaffolds of PLA reinforced with Zn particles by FFF has been demonstrated for the first time. To this end, filaments of 2.85 mm in diameter of PLA reinforced with different weight fractions of $\mu$m-sized Zn - 1 wt.% Mg alloy particles (in the range 3.5 to 17.5 wt.%) were manufactured by a double extrusion method in which standard extrusion is followed by a precision extrusion in a filament-maker machine. Filaments with constant diameter, negligible porosity and a homogeneous reinforcement distribution were obtained for Zn weight fractions of up to 10.5%. It is found that the presence of Zn particles led to changes in the physico-chemical properties of the PLA. In particular, Zn particles favored the nucleation of crystals and reduced $T_{cc}$. Moreover, the start and end degradation temperatures and the average MW of the PLA are reduced with the Zn content after precision extrusion at 200 °C. Nevertheless, all these changes were limited and did not affect the window temperature for 3D printing by FFF nor lead to significant changes in the MFI. Thus, it was possible to print scaffolds with complex shape at 190 °C with PLA/Zn composites containing 3.5 and 7 wt.% of Zn and at 170 °C when the Zn content was 10.5 wt.%. Composites with higher Zn content could not be printed because of clogging of the nozzle. The dimensional accuracy of FCC scaffolds of 20 × 20 × 20 mm$^3$ was excellent and the porosity was negligible in scaffolds with 7 wt.% of Zn content, very likely because of the viscosity reduction that facilitates 3D printing. The mechanical properties of the scaffolds containing 7 wt.% of Zn were similar to those of the PLA scaffolds in terms of stiffness and strength, although the presence of Zn particles led to a more brittle failure mode.

### CRediT authorship contribution statement

**C. Pascual-González:** Investigation, Writing – original draft, Writing – review & editing. **J. de la Vega:** Writing – review & editing, Investigation. **C. Thompson:** Investigation, Writing – review & editing. **J.P. Fernández-Blázquez:** Writing – review & editing, Methodology, Investigation. **D. Herráez-Molinero:** Investigation. **N. Biurrun:** Investigation. **I. Lizarralde:** Investigation. **J. Sánchez del Río:** Writing – review & editing, Investigation. **C. González:** Conceptualization, Investigation, Methodology, Supervision, Writing – review & editing. **J. LLorca:** Writing – review & editing, Writing – original draft, Supervision, Project administration, Methodology, Investigation, Funding acquisition, Conceptualization.

### Declaration of competing interest

The authors declare that they have no known competing financial interests or personal relationships that could have appeared to influence the work reported in this paper.

### Acknowledgments

This investigation was supported by the European Unions Horizon 2020 research and innovation programme under the European Training Network BioImpant (Development of improved bioresorbable materials for orthopaedic and vascular implant applications), Marie Sklodowska-Curie grant agreement 813869. Additional support from the Spanish Ministry of Science and Innovation through the project ADDICOMP (grant RTI2018-094435-B-C33) is gratefully acknowledged.

**References**

Agüero, A., d. C. Morcillo, M., Quiles-Carrillo, L., Balart, R., Boronat, T., Lascano, D., Torres-Giner, S., Fenollar, O., 2019. Study of the influence of the reprocessing cycles on the final properties of polylactide pieces obtained by injection molding. Polymers 11 (12), 1908.

Ali, W., Echeverry-Rendón, M., Kopp, C.G.A., LLorca, J., 2021. Strength, corrosion resistance and cellular response of interfaces in bioresorbable poly-lactic acid/mg fiber composites for orthopedic applications. J. Mech. Behav. Biomed. Mater. 123, 104781.

Antoniac, I., Popescu, D., Zapciu, A., Antoniac, A., Miculescu, F., Moldovan, H., 2019. Magnesium filled polylactic acid (PLA) material for filament based 3D printing. Materials 12 (5), 719.

Auricchio, F., Marconi, S., 2016. 3D printing: clinical applications in orthopaedics and traumatology. EFORT Open Rev. 1 (5), 121–127.

Chen, C., Lv, G., Pan, C., Song, M., Wu, C., Guo, D., Wang, X., Chen, B., Gu, Z., 2007. Poly (lactic acid)(PLA) based nanocomposites—A novel way of drug-releasing. Biomed. Mater. 2 (4), L1.

Cifuentes, S.C., Lieblich, M., López, F., Benavente, R., González-Carrasco, J.L., 2017. Effect of Mg content on the thermal stability and mechanical behaviour of PLLA/Mg composites processed by hot extrusion. Mater. Sci. Eng.: C 72, 18–25.

Colón Quintana, J.L., Heckner, T., Chrupala, A., Pollock, J., Goris, S., Osswald, T., 2019. Experimental study of particle migration in polymer processing. Polym. Compos. 40 (6), 2165–2177.

da Silva, D., Kaduri, M., Poley, M., Adir, O., Krinsky, N., Shainsky-Roitman, J., Schroeder, A., 2018. Biocompatibility, biodegradation and excretion of polylactic acid (PLA) in medical implants and theranostic systems. Chem. Eng. J. 340, 9–14.

Echeverry-Rendón, V.D.M., Quintero, D., Harmsen, M.C., Echeverría, F., 2018. Novel coatings obtained by plasma electrolytic oxidation to improve the corrosion resistance of magnesium-based biodegradable implants. Surf. Coat. Technol. 354, 28–37.

Esmaily, M., Svensson, J., Fajardo, S., Birbilis, N., Frankel, G., Virtanen, S., Arrabal, R., Thomas, S., Johansson, L., 2017. Fundamentals and advances in magnesium alloy corrosion. Prog. Mater. Sci. 89, 92–193.

Feliu Jr., S., Pardo, A., Merino, M., Coy, A., Viejo, F., Arrabal, R., 2009. Correlation between the surface chemistry and the atmospheric corrosion of AZ31, AZ80 and AZ91D magnesium alloys. Appl. Surf. Sci. 255, 4102–4108.

Ferrández-Montero, A., Lieblich, M., Benavente, R., González-Carrasco, J.L., Ferrari, B., 2020. Study of the matrix-filler interface in PLA/Mg composites manufactured by material extrusion using a colloidal feedstock. Addit. Manuf. 33, 101142.

Ferrández-Montero, A., Lieblich, M., González-Carrasco, J.L., Benavente, R., Lorenzo, V., Detsch, R., Boccaccini, A., Ferrari, B., 2019. Development of biocompatible and fully bioabsorbable PLA/Mg films for tissue regeneration applications. Acta Biomater. 98, 114–124.

Fraga, C.G., 2005. Relevance, essentiality and toxicity of trace elements in human health. Mol. Aspects Med. 26, 235–244.

Gong, H., Wang, K., Strich, R., Zhou, J.G., 2015. In vitro biodegradation behavior, mechanical properties, and cytotoxicity of biodegradable Zn-Mg alloy. J. Biomed. Mater. Res. Part B 103, 1632–1640.

Goris, S., Osswald, T., 2018. Process-induced fiber matrix separation in long fiber-reinforced thermoplastics. Composites A 105, 321–333.

Gunatillake, P.A., Adhikari, R., 2003. Biodegradable synthetic polymers for tissue engineering. Eur. Cell Mater. 5 (1), 1–16.

Hennig, B., Slim, R., Toborek, M., Robertson, L.W., 1999. Linoleic acid amplifies polychlorinated biphenyl-mediated dysfunction of endothelial cells. J. Biochem. Mol. Toxicol. 13, 83–91.

Jiang, D., Ning, F., 2020. Fused filament fabrication of biodegradable PLA/316L composite scaffolds: Effects of metal particle content. Procedia Manuf. 48, 755–762.

Kopp, A., Derra, T., Müther, M., Jauer, L., Schleifenbaum, J.H., Voshage, M., Jung, O., Smeets, R., Kröger, N., 2019. Influence of design and postprocessing parameters on the degradation behavior and mechanical properties of additively manufactured magnesium scaffolds. Acta Biomater. 98, 23–35.

Li, M., Benn, F., Derra, T., Kroeger, N., Zinser, M., Smeets, R., Molina-Aldareguía, J.M., Kopp, A., LLorca, J., 2021. Microstructure, mechanical properties, corrosion resistance and cytocompatibility of WE43 Mg alloy scaffolds fabricated by laser powder bed fusion for biomedical applications. Mater. Sci. Eng. C 119, 111623.

Li, C., Guo, C., Fitzpatrick, V., Ibrahim, A., Zwierstra, M.J., Hanna, P., Letchtig, A., Lin, S.J., Kaplan, D.L., 2020. Design of biodegradable, implantable devices towards clinical transition. Nature Rev. Mater. 5, 61–81.

Liu, S., Esteban-Manzanares, G., LLorca, J., 2020. First-principles analysis of precipitation in Mg-Zn alloys. Phys. Rev. Mater. 4, 093609.

Maurus, P.B., Kaeding, C.C., 2004. Bioabsorbable implant material review. Oper. Tech. Sports Med. 12 (3), 158–160.

Motoyama, T., Tsukegi, T., Shirai, Y., Nishida, H., Endo, T., 2007. Effects of MgO catalyst on depolymerization of poly-L-lactic acid to L, L-lactide. Polym. Degrad. Stab. 92, 1350–1358.

Murni, N.S., Dambatta, M.S., Yeap, S.K., Froemming, G.R.A., Hermawan, H., 2015. Icytotoxicity evaluation of biodegradable Zn-3Mg alloy toward normal human osteoblast cells. Mater. Sci. Eng. C 49, 560–566.

Pascual-González, C., Thompson, C., de la Vega, J., Biurrun Churruca, N., Fernández-Blázquez, J.P., Lizarralde, I., Herráez-Molinero, D., González, C., LLorca, J., 2022. Processing and properties of PLA/Mg filaments for 3D printing of scaffolds for biomedical applications. Rapid Prototyp. J. 28, 884–894.

Rokkanen, P.U., Böstman, O., Hirvensalo, E., Mäkelä, E.A., Partio, E.K., Pätiälä, H., Vainionpää, S., Vihtonen, K., Törmälä, P., 2000. Bioabsorbable fixation in orthopaedic surgery and traumatology. Biomaterials 21, 2607–2613.

Sheikh, Z., Najeeb, S., Khurshid, Z., Verma, V., Rashid, H., Glogauer, M., 2015. Biodegradable materials for bone repair and tissue engineering applications. Materials 8 (9), 5744–5794.

Shen, C., Liu, X., Fan, B., Lan, P., Zhou, F., Li, X., Wang, H., Xiao, X., Li, L., Zhao, S., 2016. Mechanical properties, in vitro degradation behavior, hemocompatibility and cytotoxicity evaluation of Zn-1.2 Mg alloy for biodegradable implants. RSC Adv. 6, 86410–86419.

Singh, R., Bathaei, M.J., Istifa, E., Beker, L., 2020. A review of bioresorbable implantable medical devices: materials, fabrication and implementation. Adv. Healthc. Mater. 9, 2000790.

Tyler, B., Gullotti, D., Mangraviti, A., Utsuki, T., Brem, H., 2016. Polylactic acid (PLA) controlled delivery carriers for biomedical applications. Adv. Drug Deliv. Rev. 107, 163–175.

Wang, S., Capoen, L., D'hooge, D.R., Cardon, L., 2018. Can the melt flow index be used to predict the success of fused deposition modelling of commercial poly (lactic acid) filaments into 3D printed materials? Plast. Rubber Compos. 47 (1), 9–16.